# Observation of light emission from Hamamatsu R11410-20 photomultiplier tubes

D.Yu. Akimov [a,b], A.I. Bolozdynya [a], Yu.V. Efremenko [a,c], V.A. Kaplin [a], A.V. Khromov [a], Yu.A. Melikyan [a], V.V. Sosnovtsev [a]

[a] *The Laboratory for Experimental Nuclear Physics of the National Research Nuclear University MEPhI (Moscow Engineering Physics Institute),*
*31 Kashirskoe Hwy, Moscow, 115409, Russian Federation*

[b] *SSC RF Institute for Theoretical and Experimental Physics of National Research Centre "Kurchatov Institute", 25 Bolshaya Cheremushkinskaya, Moscow, 117218, Russian Federation*

[c] *University of Tennessee, 1408 Circle Dr, Knoxville, TN 37996-1200, U.S.A.*

*Corresponding author: Yury Melikyan, velyour@yandex.ru; NRNU MEPhI, 31 Kashirskoe Hwy, 115409, Russian Federation; phone +7-495-788-5699 \*9015*

**Abstract** – We have shown that high voltage biased Hamamatsu R11410-20 photomultipliers with a dark count rate above 10 kHz emit single photons. The effect has been observed in a few units at room temperature and temperatures reduced down to -60 degrees Celsius. The effect should be taken into account in experiments aimed on search for rare events with small energy depositions in massive liquid xenon detectors.

The Hamamatsu R11410 photomultiplier tube (PMT) is a perspective VUV light photodetector specially designed for low-background experiments based on a liquid xenon detector technology [1-4]. The Hamamatsu R11410-20 PMT body is made of cobalt free Kovar metal that allows the operation of the PMT in the Earth's magnetic field without additional magnetic shielding. The low radioactive PMT is equipped with 76 mm diameter synthetic silica window and 64 mm diameter bialkali photocathode with quantum efficiency of about 30% at 175 nm. The typical bias voltage is ~ 1500V, with a maximum value of 1750V.

Thirty four R11410-20 PMTs were purchased and then characterized in the Laboratory for Experimental Nuclear Physics of NRNU MEPhI to be installed into the RED100 neutrino detector [5]. The majority of PMTs demonstrated dark count rate between 0.5 and 8 kHz at room temperature at the same gain of $8*10^6$. Two PMTs from this set have shown the abnormally high and unstable dark count rate: 20 kHz and 60 kHz (KB0054 and KB0018 serial numbers, respectively).

In this study, we have checked a hypothesis that abnormally high dark rate may be associated with light emission from the PMT dynode system. The experimental setup included two independently biased PMT units situated "face-to-face" inside a light-tight box. There was an optical insulation between PMT windows that could be installed or removed. The box could be cooled down with a rate of 0.2 K/min.

At room temperature, the dark count rates have been measured for the previously selected noisy KB0054 PMT and for the good one (KB0019 serial number; dark count rate <1 kHz). The KB0054 PMT bias voltage $U_{54}$ was variable between 1200 V and 1750 V with 50 V steps. The KB0019 PMT bias voltage was maintained constant at 1600V. Before testing, the both PMTs were biased and stored in darkness for at least 2 hours. The dark count rate was measured with a threshold of 1/3 of a photoelectron in dependence on applied $U_{54}$ voltage. The measured dependencies of the optically coupled PMTs on $U_{54}$ are shown in Fig.1.

As one can see, the count rate of the KB0019 PMT is proportional to that of the KB0054 PMT at $U_{54} > 1350V$. To check the assumption that the effect is associated with light emission from the KB0054 PMT, the second set of measurements was performed with optically decoupled KB0019 and KB0054 PMTs as shown in Fig.2.

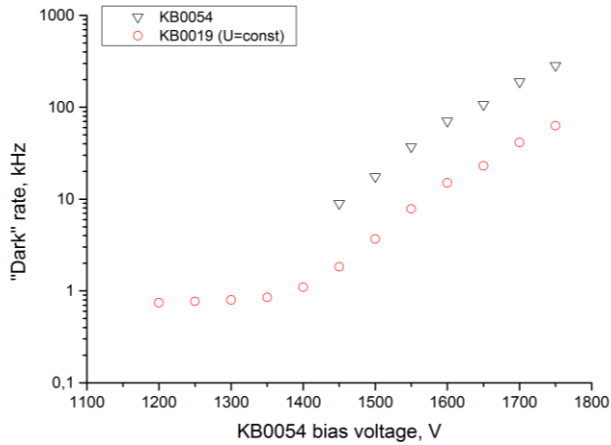

Fig.1. Dark count rates of optically coupled KB0019 and KB0054 PMTs versus $U_{54}$ voltage.

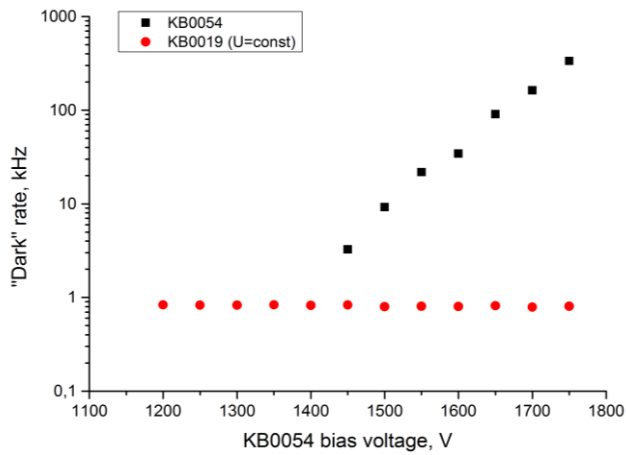

Fig.2. Dark count rates of optically decoupled KB0019 and KB0054 PMTs versus $U_{54}$ voltage.

The dramatic difference in behavior of dependencies for the KB0019 PMT in Fig.1 and Fig.2 may serve as evidence of light emission from the noisy KB0054 PMT. It is important to note that the amplitude spectrum of the dark counts from the KB0054 PMT behaves as a typical single photoelectron spectrum. No time coincidence between dark pulses detected from optically coupled KB0019 and KB0054 PMTs has been observed. This also indicates that the light emission is associated with random emission of single photons.

The next test was performed with optically coupled KB0021 and KB0019 PMTs being cooled down to −60°C. The both PMTs were characterized with low dark count rate (<1 kHz) at room temperature but they behaved differently at low temperatures. In Fig.3 the following dependencies are presented:

- dark count rate of the KB0021 PMT versus temperature when the KB0019 PMT is turned off (squares);
- dark count rate of the KB0021 PMT versus temperature when the KB0019 PMT is turned on (circles);
- dark count rate of the KB0019 PMT versus temperature when the KB0021 PMT is turned off (triangles picked up);
- dark count rate of the KB0019 PMT versus temperature when the KB0021 PMT is turned on (triangles picked down).

As seen in Fig.3, after cooling the KB0019 PMT to −45°C and below, its dark count rate raised up dramatically regardless of the KB0021 PMT operational status. At the same time, the dark count rate of the KB0021 PMT remained small (<100 Hz) at low temperatures while the KB0019 PMT was off, and rose up significantly when the KB0019 PMT was turned on.

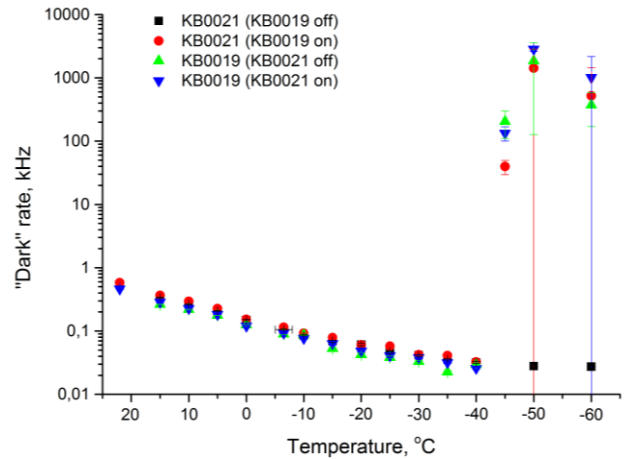

Fig.3. The dark rate versus temperature for KB0019 and KB0021 PMTs.

Finally, we have cross-checked the results of the two sets of experiments described above and have measured the dark count rate of the KB0054 PMT optically coupled to the KB0021 PMT versus $U_{54}$ at −60°C (Fig.4). The result is consistent with the result shown in Fig.1.

The acquired data prove our suggestion that the high frequency dark count rate observed for Hamamatsu R11410-20 PMTs at room temperature and at temperatures below −45°C is associated with the single photon emission from the PMT internal structure. Note this kind of light emission was also reported for the Hamamatsu R11065-10 PMT [6].

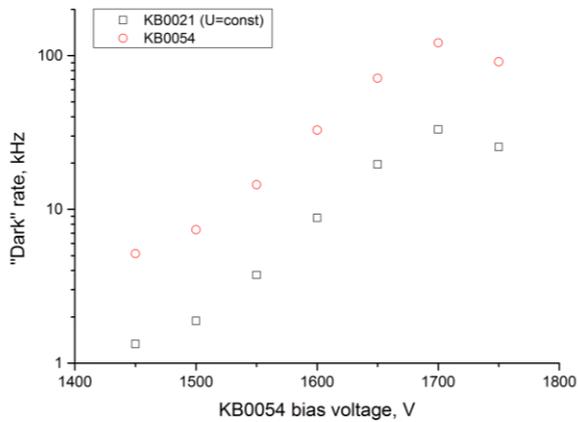

Fig.4. Dark count rates of optically coupled KB0021 and KB0054 PMTs versus $U_{54}$ measured at −60°C.

We believe that the photon emission effect can be observed with different level of intensity in any high-voltage biased PMT. The effect may depend on construction materials, temperature and design of a PMT. The effect cannot essentially affect registration of multi-photon signals with several PMTs operated in the coincidence mode because of random character of the single photon emission. However, detection of weak scintillation signals with a large number of PMTs viewing the same detection volume (such as in the LZ detector [7]) could be affected by this effect. For this type of the detector setup PMTs should be pre-selected on relatively low (<1 kHz) dark count rate in the temperature range of operation.

**Acknowledgment**

This study was supported by the RF Government under the contract of the NRNU MEPhI with the Ministry of Education and Science of №11.G34.31.0049 from October 19, 2011.